\documentclass[11pt,a4paper,fleqn]{article}

\usepackage{amsmath,amssymb,graphicx,siunitx,multirow,float,calc,array,tabularx,caption,float,lipsum,subcaption}
\usepackage[table]{xcolor}
\usepackage[colorlinks=true,breaklinks=true,allcolors=blue]{hyperref}
\newcommand{\revision}[1]{{#1}}
\usepackage[numbers,sort&compress]{natbib}
\usepackage{authblk}
\usepackage[margin=1.9cm]{geometry}

\begin{document}
\title{Analytical formulas for calculating the thermal diffusivity of cylindrical shell and spherical shell samples}
\author{{Elliot J. Carr\thanks{\href{mailto:elliot.carr@qut.edu.au}{elliot.carr@qut.edu.au}}\,\,} and Luke P. Filippini}
\affil{School of Mathematical Sciences, Queensland University of Technology, Brisbane, Australia.}

\date{}

\maketitle

\noindent\rule{\textwidth}{0.5pt}\\[-1.0cm]
\section*{Abstract}
Calculating the thermal diffusivity of solid materials is commonly carried out using the laser flash experiment. This classical experiment considers a small (usually thin disc-shaped) sample of the material with parallel front and rear surfaces, applying a heat pulse to the front surface and recording the resulting rise in temperature over time on the rear surface. Recently, Carr and Wood [Int J Heat Mass Transf, 144 (2019) 118609] showed that the thermal diffusivity can be expressed analytically in terms of the heat flux function applied at the front surface and the temperature rise history at the rear surface. In this paper, we generalise this result to radial unidirectional heat flow, developing new analytical formulas for calculating the thermal diffusivity for cylindrical shell and spherical shell shaped samples. Two configurations are considered: (i) heat pulse applied on the inner surface and temperature rise recorded on the outer surface and (ii) heat pulse applied on the outer surface and temperature rise recorded on the inner surface. Code implementing and verifying the thermal diffusivity formulas for both configurations is made available.\\
\\
\textbf{Keywords:} laser flash; thermal diffusivity; parameter estimation; heat flow; cylindrical shell; spherical shell.

\noindent\rule{\textwidth}{0.5pt}

\section{Introduction}
Thermal diffusivity is an important property measuring the rate at which heat transfers through a material. The most popular way to measure the thermal diffusivity of solid materials is to perform the laser flash experiment, which was was first developed by \citet{parker_1961} for the case of a homogeneous, isotropic, thermally insulated, disc-shaped slab. Several alternative approaches have since been proposed to calculate the thermal diffusivity from the laser flash experiment. These techniques differ either in the mathematical model used to describe the temperature distribution within the sample over time or the mathematical/computational method used to fit the theoretical rear-surface temperature rise profile to experimental data.

Modifications to the mathematical model have been made to accommodate additional physical effects such as the shape and duration of the heat pulse \cite{cape_1963,azumi_1981,carr_2019d}, heat loss between the sample and the environment \cite{parker_1962,chen_2010,carr_2019b} and layered samples \revision{\cite{larson_1968,czel_2013,carr_2019d,milosevic_2004}}. Besides the half-rise time approach, other data reduction methods include the logarithmic method \cite{takahashi_1988,thermitus_1997,chihab_2020,nishi_2020}, the ratio method \cite{clark_1975,gosset_2002,vozar_2003} and the moment method \cite{degiovanni_1986,vozar_2003}, each using analytical forms of the theoretical rear-surface temperature rise profiles to develop \textit{approximate} analytical formulas for the thermal diffusivity. Nonlinear least-squares curve fitting is also commonly used to calculate the thermal diffusivity numerically, where the aim is to minimise the sum of squared differences between the experimental and theoretical rear-surface temperature rise profiles at each experimentally sampled point in time \revision{\cite{chen_2010,czel_2013,gembarovic_1990,el-rassy_2020,lunev_2020,milosevic_2019}}. Numerical solutions of the governing heat flow model have also been combined with Bayesian methods \cite{lamien_2019} and neural network techniques \cite{yan_2022} to compute the thermal diffusivity numerically. 

\begin{figure}[t]
\hypertarget{fig:cylindrical_shell}{}
\centering
\fbox{\includegraphics[width=0.42\textwidth,trim=3cm 2.7cm 3cm 3cm,clip]{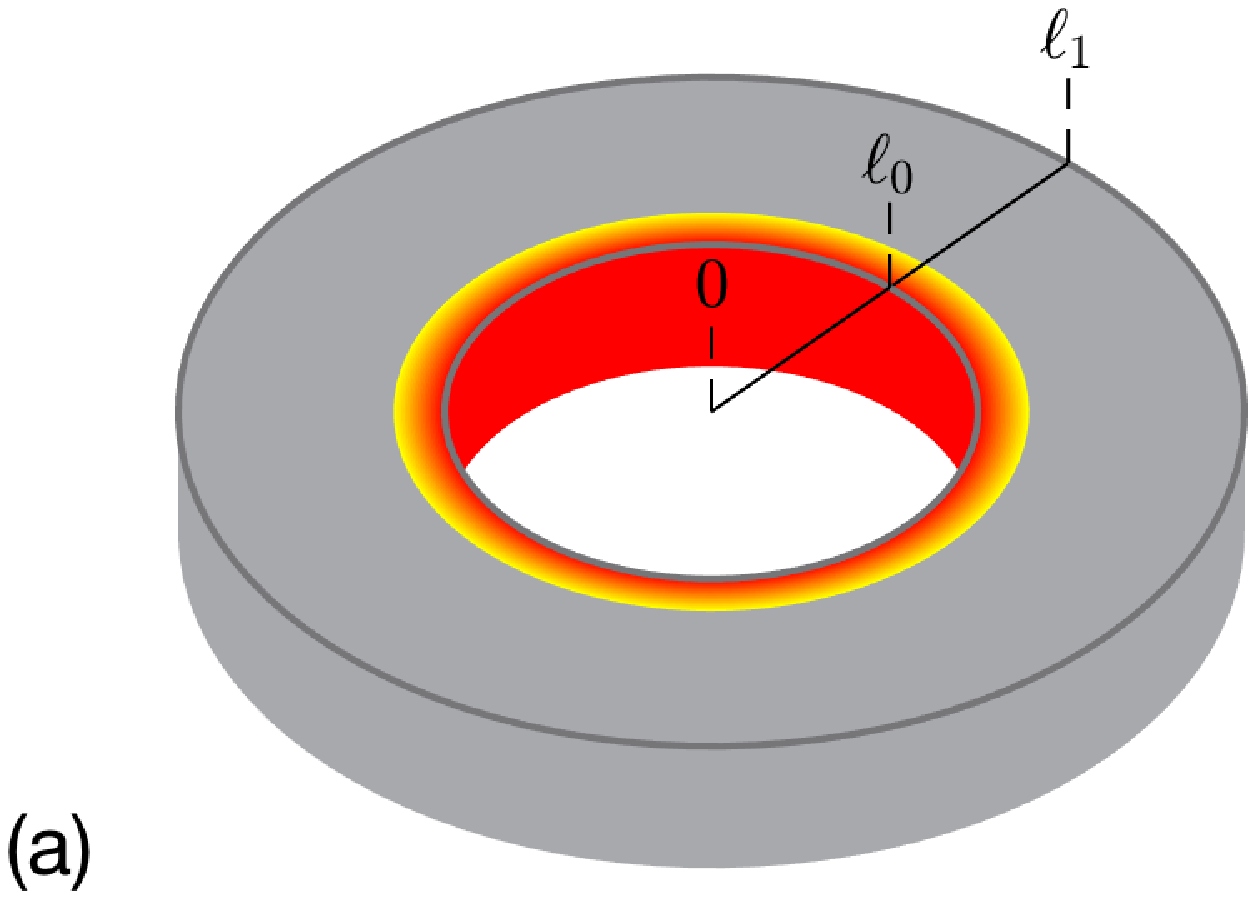}\hspace{1.3cm}
\includegraphics[width=0.42\textwidth,trim=3cm 2.7cm 3cm 3cm,clip]{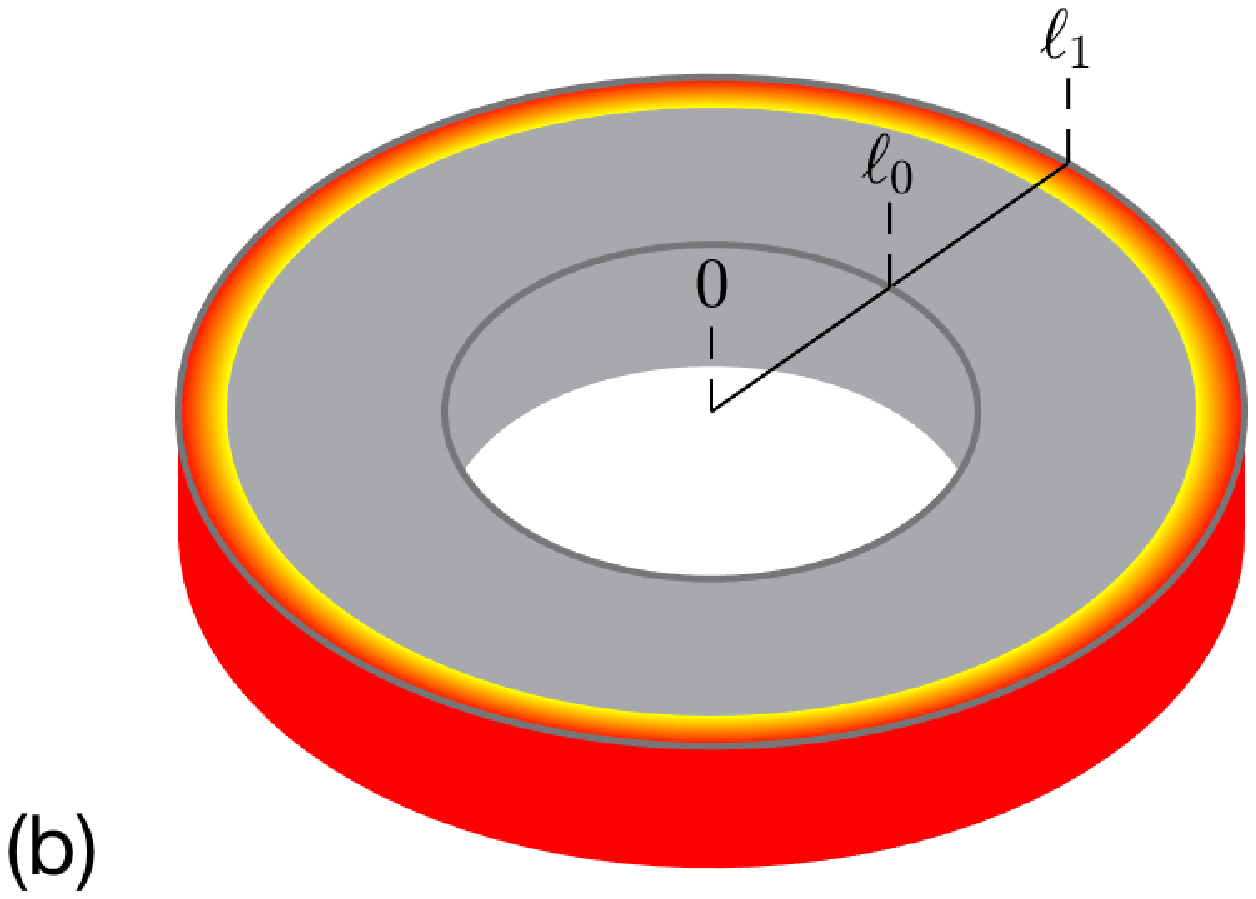}}\\
\caption{\textbf{Cylindrical shell shaped samples}. (a) outward configuration where the heat pulse is applied uniformly on the inner surface ($x=\ell_{0}$) and the temperature rise is recorded on the outer surface ($x=\ell_{1}$) and (b) inward configuration where the heat pulse is applied uniformly on the outer surface ($x=\ell_{1}$) and the temperature rise is recorded on the inner surface ($x=\ell_{0}$).}
\end{figure}

\begin{figure}[t]
\hypertarget{fig:spherical_shell}{}
\centering
\fbox{\includegraphics[width=0.42\textwidth,trim=3cm 2.7cm 3cm 1.5cm,clip]{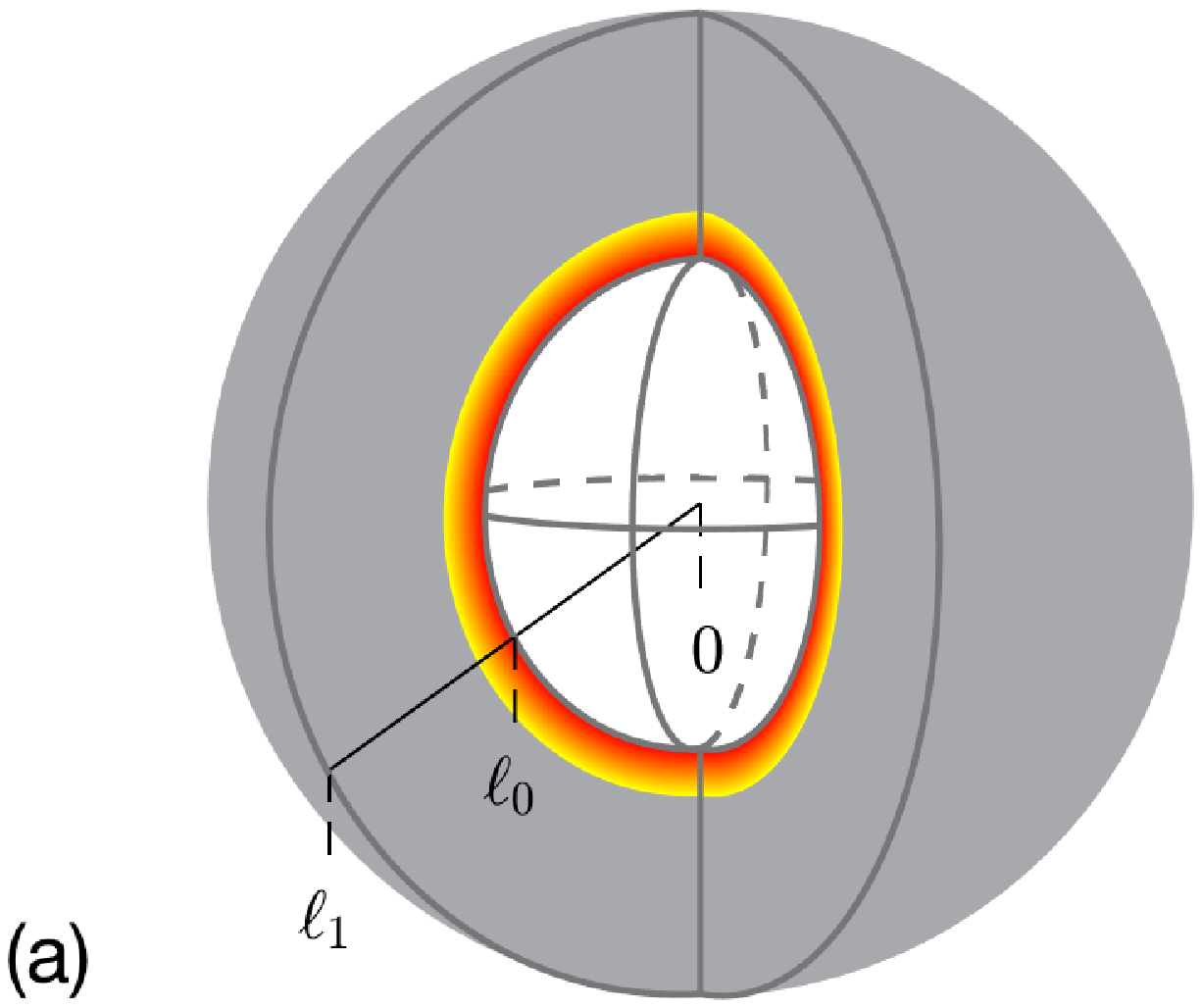}\hspace{1.3cm}
\includegraphics[width=0.42\textwidth,trim=3cm 2.7cm 3cm 1.5cm,clip]{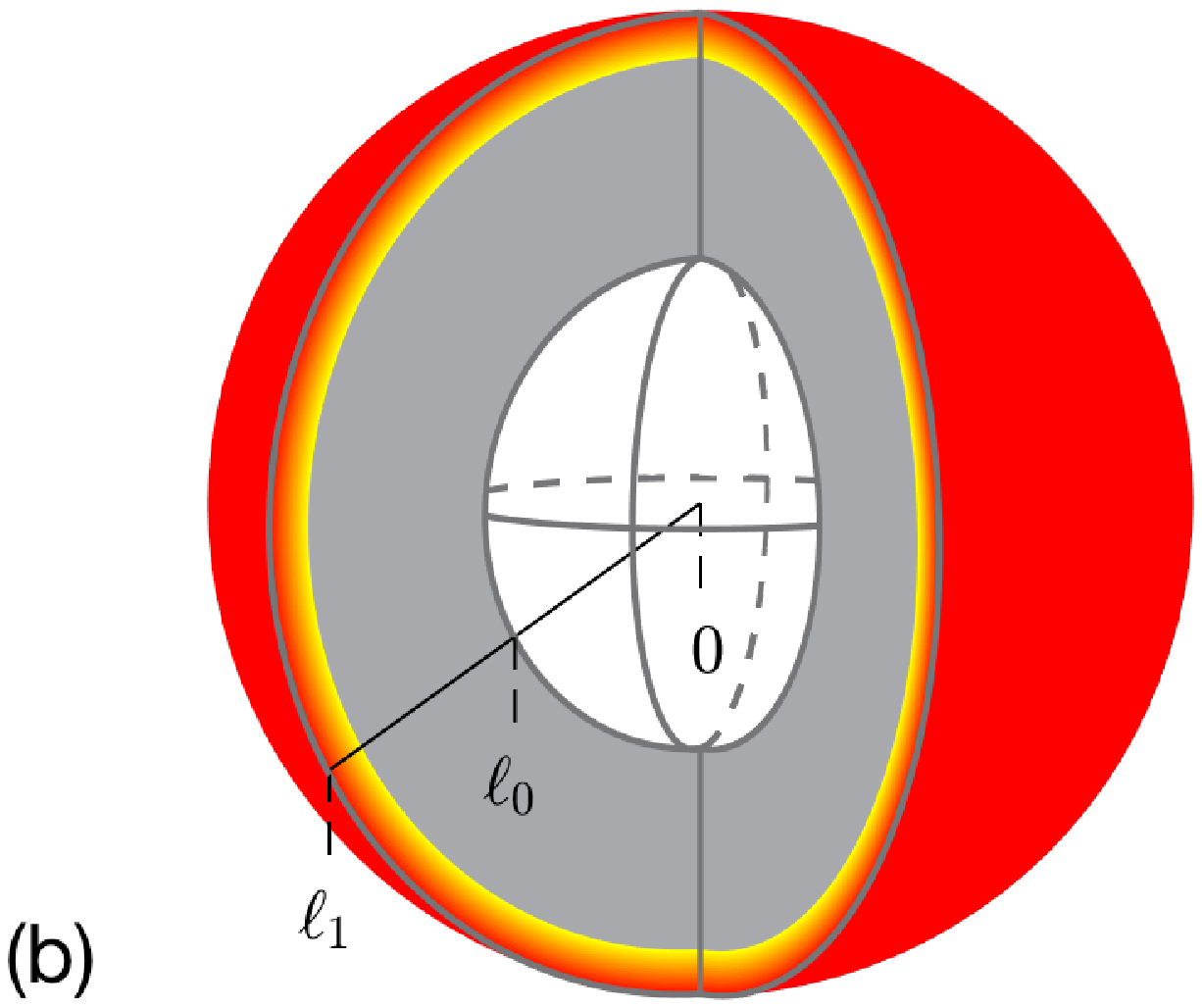}}
\caption{\textbf{Spherical shell shaped samples}. (a) outward configuration where the heat pulse is applied uniformly on the inner surface ($x=\ell_{0}$) and the temperature rise is recorded on the outer surface ($x=\ell_{1}$) and (b) inward configuration where the heat pulse is applied uniformly on the outer surface ($x=\ell_{1}$) and the temperature rise is recorded on the inner surface ($x=\ell_{0}$).}
\end{figure}

In this work, we focus on data reduction methods that explicitly and exactly express the thermal diffusivity in terms of the theoretical rear-surface temperature rise profile. Such data reduction methods include the \textit{areal heat diffusion method} \cite{baba_2009} and the \textit{rear-surface integral method} \cite{carr_2019b,carr_2019d}. Both of these methods develop thermal diffusivity formulas for a homogeneous, isotropic, thermally insulated, disc-shaped slab with parallel front and rear surfaces, assuming uniform heating at the front surface and uniform initial temperature. Under these conditions, if $\mathcal{T}(x,t)$ is the temperature of the sample at location $x$ and time $t$ then the \textit{temperature rise}, $T(x,t) = \mathcal{T}(x,t) - \mathcal{T}_{0}$ above the uniform initial temperature $\mathcal{T}_{0}$, satisfies the heat flow model:
\begin{gather}
\label{eq:pde_slab}
\frac{\partial T}{\partial t} = \alpha\frac{\partial^{2} T}{\partial x^{2}},\quad 0 < x < L,\quad t > 0,\\
\label{eq:ic_slab}
T(x,0) = 0,\quad 0 < x < L,\\
\label{eq:bcs_slab}
-k\frac{\partial T}{\partial x}(0,t) = q(t),\quad\frac{\partial T}{\partial x}(L,t) = 0,
\end{gather}
where $q(t)$ is the heat flux corresponding to the heat pulse applied at the front surface ($x=0$). Let $Q(t) = \int_{0}^{t}q(s)\,\text{d}s$, $Q_{\infty} = \lim_{t\rightarrow\infty} Q(t) = \int_{0}^{\infty}q(t)\,\text{d}t$ and $T_{\infty} = \lim_{t\rightarrow\infty}T(x,t)$. \citet{baba_2009} derived the following formula for the thermal diffusivity:
\begin{gather}
\label{eq:alpha_baba}
\alpha = \frac{L^{2}}{6\int_{0}^{\infty}[1 - T(L,t)/T_{\infty}]\,\text{d}t},
\end{gather}
assuming a perfectly instantaneous heat pulse, $q(t) = Q_{\infty}\delta(t)$ where $\delta(t)$ is the Dirac delta function. \citet{carr_2019d} later generalising (\ref{eq:alpha_baba}) to an arbitrary heat pulse function: 
\begin{gather}
\label{eq:alpha_carr2019d}
\alpha = \frac{L^{2}}{6(\int_{0}^{\infty}[1 - T(L,t)/T_{\infty}]\,\text{d}t-\int_{0}^{\infty}[1 - Q(t)/Q_{\infty}]\,\text{d}t)},
\end{gather}
assuming only that $\lim_{t\rightarrow\infty}q(t) = 0$ (the heat pulse is eventually switched off) and $Q_{\infty}$ is finite (the total amount of heat absorbed into the sample is finite). Both formulas express the thermal diffusivity exactly in terms of the theoretical rear-surface temperature rise curve, $T(L,t)$, obtained from the heat flow model (\ref{eq:pde_slab})--(\ref{eq:bcs_slab}).

In this paper, motivated by the laser flash \revision{experiment \cite{parker_1961}}, we further build on the work presented in \cite{carr_2019b,carr_2019d} by generalising (\ref{eq:alpha_carr2019d}) to radial unidirectional heat flow in cylindrical shell and spherical shell shaped samples as shown in \hyperlink{fig:cylindrical_shell}{Figure 1} and \hyperlink{fig:spherical_shell}{Figure 2}, respectively. In such samples, $x\in[\ell_{0},\ell_{1}]$ is now the radial coordinate and the direction of the flow of heat (inward towards $x = \ell_{0}$ or outward towards $x = \ell_{1}$) becomes important, leading to two distinct cases: \textit{outward configuration}, where the heat pulse is applied on the inner surface at $x = \ell_{0}$ and the temperature rise is recorded on the outer surface at $x = \ell_{1}$ (\hyperlink{fig:cylindrical_shell}{Figures 1(a) and 2(a)}) and \textit{inward configuration}, where the heat pulse is applied on the outer surface at $x = \ell_{1}$ and the temperature rise is recorded on the inner surface at $x = \ell_{0}$ (\hyperlink{fig:cylindrical_shell}{Figures 1(b) and 2(b)}). While we acknowledge that formulas for calculating the thermal diffusivity of such geometries may be of limited practical use in real laser flash analysis, the new formulas increase the fundamental understanding of how the heat or mass diffusion equation (\ref{eq:pde_slab}) can be parameterised using boundary data, revealing the role of dimension $d$ in the calculations ($d=1,2,3$ for disc-shaped slab, cylindrical shell and spherical shell, respectively) and highlighting how little effect the direction of the flow of heat actually has on the resulting formula. 

\hypertarget{sec:analytical_formula}{}
\section{Analytical formula}
\hypertarget{sec:outward_configuration}{}
\subsection{Outward configuration}
We first consider developing a formula for the thermal diffusivity, $\alpha$, for the case where the heat pulse is applied at the inner surface ($x = \ell_{0}$) and the temperature rise is recorded at the outer-surface ($x=\ell_{1}$). In this case, the change in temperature within the sample, $T(x,t)$, satisfies the $d$-dimensional unidirectional generalisation of the heat flow model (\ref{eq:pde_slab})--(\ref{eq:bcs_slab}):
\begin{gather}
\label{eq:pde_outward}
\frac{\partial T}{\partial t} = \frac{\alpha}{x^{d-1}}\frac{\partial}{\partial x}\left(x^{d-1}\frac{\partial T}{\partial x}\right),\quad \ell_{0} < x < \ell_{1},\quad t > 0,\\
\label{eq:ic_outward}
T(x,0) = 0,\quad \ell_{0} < x < \ell_{1},\\
\label{eq:bcs_outward}
-k\frac{\partial T}{\partial x}(\ell_{0},t) = q(t),\quad\frac{\partial T}{\partial x}(\ell_{1},t) = 0,
\end{gather} 
where $d=2$ corresponds to the cylindrical shell (\hyperlink{fig:cylindrical_shell}{Figure 1(a)}) and $d = 3$ corresponds to the spherical shell (\hyperlink{fig:spherical_shell}{Figure 2(a)}). Before developing our new formulas for the thermal diffusivity, we present two important results.

\textit{Conservation of heat:} Conservation of heat requires that the change in heat within the cylindrical/spherical shell, $\Omega_{d}$, is balanced by the heat entering through the inner surface, $\Gamma_{0}$, at $x = \ell_{0}$:
\begin{gather*}
\rho c\int_{\Omega_{d}} T(x,t)\,\text{d}V = \int_{\Gamma_{0}}Q(t)\,\text{d}A.
\end{gather*}
Due to radial symmetry we have:
\begin{gather*}
\rho c\int_{\Omega_{d}} T(x,t)\,\text{d}V = \rho c S_{d}\int_{\ell_{0}}^{\ell_{1}} x^{d-1}T(x,t)\,\text{d}x\quad\text{and}\quad\int_{\Gamma_{0}}Q(t)\,\text{d}A = S_{d}\ell_{0}^{d-1}Q(t), 
\end{gather*}
where $S_{d}$ is the surface area of the $d$-dimensional unit sphere. Hence, conservation of heat requires 
\begin{gather}
\label{eq:heat_conservation}
\rho c\int_{\ell_{0}}^{\ell_{1}} x^{d-1}T(x,t)\,\text{d}x = \ell_{0}^{d-1}Q(t),
\end{gather}
for all $t > 0$. 

\textit{Steady state solution:} The steady state solution of the heat flow model (\ref{eq:pde_outward})--(\ref{eq:bcs_outward}), \smash{$T_{\infty} = \lim_{t\rightarrow\infty}T(x,t)$}, satisfies the boundary value problem:
\begin{gather*}
0 = \frac{\text{d}}{\text{d}x}\left(x^{d-1}\frac{\text{d}T_{\infty}}{\text{d}x}\right),\quad \ell_{0} < x < \ell_{1},\\
\frac{\text{d}T_{\infty}}{\text{d}x}(\ell_{0}) = 0,\quad\frac{\text{d}T_{\infty}}{\text{d}x}(\ell_{1}) = 0,
\end{gather*}
with the boundary condition at $x = \ell_{0}$ applying since $\lim_{t\rightarrow\infty}q(t) = 0$. Solving this differential equation for $T_{\infty}$ and applying the boundary conditions it is clear that $T_{\infty}$ is constant for all $x\in [\ell_{0},\ell_{1}]$. The constant value of $T_{\infty}$ is then identified by applying (\ref{eq:heat_conservation}) in the limit $t\rightarrow\infty$ and rearranging to give:
\begin{gather}
\label{eq:Tinf_outward}
T_{\infty} = \frac{d\ell_{0}^{d-1}Q_{\infty}}{\rho c (\ell_{1}^{d}-\ell_{0}^{d})},\quad \ell_{0} < x < \ell_{1}.
\end{gather}

We now derive our new formula for calculating the thermal diffusivity, $\alpha$. This is achieved by formulating and solving a boundary value problem satisfied by the following function \cite{carr_2019b,carr_2019d}:
\begin{gather}
\label{eq:u_definition}
u(x) = \int_{0}^{\infty} \left[T_{\infty} - T(x,t)\right]\,\text{d}t.
\end{gather}
Applying the linear operator 
\begin{gather*}
\mathcal{L} = \frac{\alpha}{x^{d-1}}\frac{\partial}{\partial x}\left(x^{d-1}\frac{\partial}{\partial x}\right),
\end{gather*}
to both sides of equation (\ref{eq:u_definition}) and using the heat equation (\ref{eq:pde_outward}), initial condition (\ref{eq:ic_outward}) and the fact that $T_{\infty}$ (\ref{eq:Tinf_outward}) is a constant yields the following differential equation satisfied by $u(x)$:
\begin{gather}
\label{eq:u_ode}
\frac{\alpha}{x^{d-1}}\frac{\text{d}}{\text{d}x}\left(x^{d-1}\frac{\text{d}u}{\text{d}x}\right) = -T_{\infty}.
\end{gather}
Solving for $u(x)$ gives
\begin{gather}
\label{eq:u_polynomial}
u(x) = c_{0} + c_{1}\int_{\ell_{0}}^{x} s^{d-1}\,\text{d}s - \frac{T_{\infty}x^{2}}{2\alpha d},
\end{gather}
where $c_{0}$ and $c_{1}$ are (as yet) undetermined integration constants. Combining the boundary conditions (\ref{eq:bcs_outward}) with the derivative
\begin{align*}
\frac{\text{d}u}{\text{d}x} = \int_{0}^{\infty} -\frac{\partial T}{\partial x}(x,t)\,\text{d}t,
\end{align*}
yields the boundary conditions satisfied by $u(x)$ at the inner and outer surfaces \cite{carr_2019d}:
\begin{align}
\label{eq:u_bcs}
\frac{\text{d}u}{\text{d}x}(\ell_{0}) = \frac{Q_{\infty}}{k},\quad
\frac{\text{d}u}{\text{d}x}(\ell_{1}) = 0.
\end{align}
Noting from (\ref{eq:u_polynomial}) that
\begin{align*}
\frac{\text{d}u}{\text{d}x} = c_{1}x^{d-1} - \frac{T_{\infty}x}{\alpha d},
\end{align*}
and using the relationship between $T_{\infty}$ and $Q_{\infty}$ (\ref{eq:Tinf_outward}) we see that both boundary conditions (\ref{eq:u_bcs}) are satisfied when
\begin{align*}
c_{1} &= \frac{T_{\infty}\ell_{1}^{d}}{\alpha d}.
\end{align*} 
To identify $c_{0}$ we require the following analogous condition to the heat conservation constraint (\ref{eq:heat_conservation})
\begin{align}
\label{eq:u_ac}
\rho c\int_{\ell_{0}}^{\ell_{1}} x^{d-1}u(x)\,\text{d}x = \ell_{0}^{d-1}\int_{0}^{\infty} \left[Q_{\infty} - Q(t)\right]\,\text{d}t,
\end{align}
which is derived by making use of equations (\ref{eq:heat_conservation}), (\ref{eq:Tinf_outward}) and (\ref{eq:u_definition}) as follows:
\begin{align*}
\rho c\int_{\ell_{0}}^{\ell_{1}}x^{d-1}u(x)\,\text{d}x &= \rho c\int_{\ell_{0}}^{\ell_{1}}x^{d-1}\int_{0}^{\infty}[T_{\infty}-T(x,t)]\,\text{d}t\,\text{d}x,\\
&=\rho c\int_{0}^{\infty}\biggl[T_{\infty}\int_{\ell_{0}}^{\ell_{1}}x^{d-1}\,\text{d}x - \int_{\ell_{0}}^{\ell_{1}}x^{d-1}T(x,t)\,\text{d}x\biggr]\text{d}t,\\
&=\ell_{0}^{d-1}\int_{0}^{\infty}[Q_{\infty} - Q(t)]\,\text{d}t.
\end{align*}
Imposing the additional condition (\ref{eq:u_ac}) on $u(x)$ (\ref{eq:u_polynomial}) identifies $c_{0}$ (not shown) and hence $u(x)$. In summary, the solution of the boundary value problem described by equations (\ref{eq:u_ode}), (\ref{eq:u_bcs}) and (\ref{eq:u_ac}) is given by
\begin{multline}
u(x) = \frac{d\ell_{0}^{d-1}\int_{0}^{\infty}[Q_{\infty} - Q(t)]\,\text{d}t}{\rho c(\ell_{1}^{d}-\ell_{0}^{d})}+ \frac{T_{\infty}}{d\alpha}\ell_{1}^{d}\int_{\ell_{0}}^{x}s^{1-d}\,\text{d}s - \frac{T_{\infty}}{2d\alpha}x^{2}\\ 
+ \frac{T_{\infty}}{\alpha(\ell_{1}^{d}-\ell_{0}^{d})}\biggl[\frac{(\ell_{1}^{d+2}-\ell_{0}^{d+2})}{2(d+2)} - \ell_{1}^{d}\int_{\ell_{0}}^{\ell_{1}}x^{d-1}\int_{\ell_{0}}^{x}s^{1-d}\,\text{d}s\,\text{d}x \biggr].\label{eq:u_exp2}
\end{multline}
As we now have a second expression for $u(x)$ (in addition to the integral expression (\ref{eq:u_definition})) equating both expressions at the outer surface, $x = \ell_{1}$, and rearranging yields a formula for the thermal diffusivity, $\alpha$, which after algebraic simplification can be expressed in the following form:
\begin{gather}
\label{eq:alpha_outward}
\alpha = \frac{\ell_{1}^{d+2}-(d+2)\ell_{0}^{d}\ell_{1}^{d}\int_{\ell_{0}}^{\ell_{1}}s^{1-d}\,\text{d}s - \ell_{0}^{d+2}}{2(d+2)(\ell_{1}^{d}-\ell_{0}^{d})[\int_{0}^{\infty}[1 - T(\ell_{1},t)/T_{\infty}]\,\text{d}t-\int_{0}^{\infty}[1 - Q(t)/Q_{\infty}]\,\text{d}t]}.
\end{gather}
This formula expresses the thermal diffusivity explicitly in terms of $T_{\infty}$, $Q_{\infty}$ and $Q(t)$; the inner and outer radii, $\ell_{0}$ and $\ell_{1}$; and the theoretical outer surface temperature rise curve, $T(\ell_{1},t)$. For the specific cases of the cylindrical shell ($d=2$) and the spherical shell ($d=3$), formula (\ref{eq:alpha_outward}) reduces~to:

\bigskip
\noindent\textit{Cylindrical shell ($d=2$)}
\begin{gather}
\label{eq:alpha2}
\alpha = \frac{\ell_{1}^{4}-4\ell_{0}^{2}\ell_{1}^{2}\log(\ell_{1}/\ell_{0}) - \ell_{0}^{4}}{8(\ell_{1}^{2}-\ell_{0}^{2})[\int_{0}^{\infty}[1 - T(\ell_{1},t)/T_{\infty}]\,\text{d}t-\int_{0}^{\infty}[1 - Q(t)/Q_{\infty}]\,\text{d}t]},
\end{gather}

\noindent\textit{Spherical shell ($d=3$)}
\begin{gather}
\label{eq:alpha3}
\alpha = \frac{\ell_{1}^{5}-5\ell_{0}^{2}\ell_{1}^{2}(\ell_{1}-\ell_{0})-\ell_{0}^{5}}{10(\ell_{1}^{3}-\ell_{0}^{3})[\int_{0}^{\infty}[1 - T(\ell_{1},t)/T_{\infty}]\,\text{d}t-\int_{0}^{\infty}[1 - Q(t)/Q_{\infty}]\,\text{d}t]},
\end{gather}
while $d = 1$ recovers formula (\ref{eq:alpha_carr2019d}) with the thickness of the sample identified as $L = \ell_{1}-\ell_{0}$. 

\hypertarget{sec:inward_configuration}{}
\subsection{Inward configuration}
For the opposite case or \text{inward configuration}, the boundary conditions (\ref{eq:bcs_outward}) are reversed yielding:
\begin{gather}
\label{eq:bcs_inward}
\frac{\partial T}{\partial x}(\ell_{0},t) = 0,\quad k\frac{\partial T}{\partial x}(\ell_{1},t) = q(t).
\end{gather}
Repeating the working of \hyperlink{sec:outward_configuration}{Section 2.1} yields a nearly identical formula for the thermal diffusivity as developed for the outward configuration (\ref{eq:alpha_outward}),
\begin{gather}
\label{eq:alpha_inward}
\alpha = \frac{\ell_{1}^{d+2}-(d+2)\ell_{0}^{d}\ell_{1}^{d}\int_{\ell_{0}}^{\ell_{1}}s^{1-d}\,\text{d}s - \ell_{0}^{d+2}}{2(d+2)(\ell_{1}^{d}-\ell_{0}^{d})[\int_{0}^{\infty}[1 - T(\ell_{0},t)/T_{\infty}]\,\text{d}t-\int_{0}^{\infty}[1 - Q(t)/Q_{\infty}]\,\text{d}t]},
\end{gather}
with the only difference being that the first integral on the denominator involves the inner surface temperature rise curve, $T(\ell_{0},t)$, instead of the outer surface temperature rise curve, $T(\ell_{1},t)$.

\section{Numerical validation}
\label{sec:results}

\hypertarget{sec:discrete_data}{}
\subsection{Application to discrete data}
The temperature rise data takes the form of a sequence of uniformly-spaced discrete-time values, $\widetilde{T}_{0},\widetilde{T}_{1},\hdots,\widetilde{T}_{N}$ where $\widetilde{T}_{i}$ is the temperature rise recorded at $t = t_{i} := i\Delta t$, with $\Delta t = t_{N}/N$ and $t_{N}$ being the time corresponding to the final recorded temperature. For the outward and inward configurations, note that $\widetilde{T}_{0},\widetilde{T}_{1},\hdots,\widetilde{T}_{N}$ are recorded at the outer ($x = \ell_{1}$) and inner ($x = \ell_{0}$) surfaces, respectively. In this case, to apply the thermal diffusivity formulas (\ref{eq:alpha_outward}) and (\ref{eq:alpha_inward}), we use the trapezoidal rule to evaluate the integral:
\begin{align}
\label{eq:trap_rule}
\int_{0}^{\infty} \left[1 - T(\ell_{0/1},t)/T_{\infty}\right]\,\text{d}t \approx \int_{0}^{t_{N}}[1 - T(\ell_{0/1},t)/T_{\infty}]\,\text{d}t \approx \Delta t\sum_{i=1}^{N} \left(1 - \frac{\widetilde{T}_{i-1}+\widetilde{T}_{i}}{2T_{\infty}}\right),
\end{align}
featuring in the denominators of (\ref{eq:alpha_outward}) and (\ref{eq:alpha_inward}), respectively. Although we consider only continuous descriptions of the heat flux $q(t)$ in this work, we note that both (\ref{eq:alpha_outward}) and (\ref{eq:alpha_inward}) can also be applied in the case of discrete descriptions of $q(t)$ by using numerical integration (e.g. the trapezoidal rule) to evaluate the integral $\int_{0}^{\infty}[1 - Q(t)/Q_{\infty}]\,\text{d}t$ featuring in the denominators of both (\ref{eq:alpha_outward}) and (\ref{eq:alpha_inward}) as discussed in \cite{carr_2019d}.

\hypertarget{sec:perfect}{}
\subsection{Application to perfect synthetic data}
We now verify our thermal diffusivity formulas (\ref{eq:alpha_outward}) and (\ref{eq:alpha_inward}) using ``perfect'' noise-free synthetic data. To generate the synthetic data we solve the governing heat flow model, equations (\ref{eq:pde_outward})--(\ref{eq:bcs_outward}) for the outward configuration and equations (\ref{eq:pde_outward}), (\ref{eq:ic_outward}) and (\ref{eq:bcs_inward}) for the inward configuration, numerically using a known set of parameter values and extract the temperature rise data at either the inner or outer surface as required. Numerical solutions of the governing heat flow model are computed by discretising in space using a finite volume method and discretising in time using MATLAB's \texttt{ode15s} solver. Full details on this numerical method are available in our supporting MATLAB code available at \href{https://github.com/elliotcarr/Carr2022c}{https://github.com/elliotcarr/Carr2022c}. This process yields the following discrete-time temperature rise data, $\widetilde{T}_{0},\widetilde{T}_{1},\hdots,\widetilde{T}_{N}$ as introduced in \hyperlink{sec:discrete_data}{Section 3.1}:
\begin{gather*}
\widetilde{T}_{i} = \begin{cases} T_{i}^{(0)}, & \text{for the inward configuration},\\
T_{i}^{(1)}, & \text{for the outward configuration}, \end{cases}
\end{gather*}
where $T_{i}^{(0)}$ and $T_{i}^{(1)}$ are the numerical approximations to $T(\ell_{0},t_{i})$ and $T(\ell_{1},t_{i})$ obtained from the governing heat flow model. In this paper, all results are reported using a commonly used parameter set \cite{czel_2013,carr_2019d}: 
\begin{gather}
\label{eq:parameters1}
k = 222\,\text{W}\,\text{m}^{-1}\text{K}^{-1},\quad\rho = 2700\,\text{kg}\,\text{m}^{-3},\quad c = 896\,\text{J}\,\text{kg}^{-1}\text{K}^{-1},\\ \ell_{0} = 0.001\,\text{m},\quad \ell_{1} = 0.003\,\text{m},\quad Q_{\infty} = 7000\,\text{J}\,\text{m}^{-2},\quad \beta = 0.001\,\text{s},\\
\label{eq:parameters2}
q(t) = \frac{Q_{\infty}t}{\beta^{2}}e^{-t/\beta},\quad N = 1000,\quad \Delta t = 10^{-4}\,\text{s},\quad t_{N} = 0.1\,\text{s},
\end{gather}
where the form of $q(t)$ describes an exponential pulse \cite{larson_1968} that reaches a peak value at $t = \beta$. For this choice of $q(t)$ we have an exact expression for the integral involving $Q(t)$ in the thermal diffusivity formulas (\ref{eq:alpha_outward}) and (\ref{eq:alpha_inward}):
\begin{gather*}
\int_{0}^{\infty} \left[1 - Q(t)/Q_{\infty}\right]\,\text{d}t  = 2\beta.
\end{gather*}
The parameters (\ref{eq:parameters1}) yield a target value of the thermal diffusivity of
\begin{gather}
\label{eq:alpha_target}
\alpha = k/(\rho c) = 9.1766\times 10^{-5}\,\text{m}^{2}\text{s}^{-1}.
\end{gather}
In \hyperlink{tab:results1}{Table 1}, we compare estimated and target values of the thermal diffusivity for each geometry (disc-shaped slab, cylindrical shell, spherical shell) and configuration (outward, inward) combination. To quantify the discrepancy we state the signed relative percentage error: $\varepsilon = (\alpha-\widetilde{\alpha})/\alpha \times 100\,(\%)$, where $\alpha$ is the target value of the thermal diffusivity (\ref{eq:alpha_target}) and $\widetilde{\alpha}$ is the estimated value of the thermal diffusivity computed using (\ref{eq:alpha_outward}) and (\ref{eq:trap_rule}) for the outward configuration and (\ref{eq:alpha_inward}) and (\ref{eq:trap_rule}) for the inward configuration. Results in \hyperlink{tab:results1}{Table 1} demonstrate that the thermal diffusivity estimates agree with the target thermal diffusivity to between four and five significant figures with a corresponding relative percentage error of less than 0.001\%. Note that the same values are recorded for $d = 1$ since the inward and outward configurations are the same for the disc-shaped slab. These results verify the derivations carried out in \hyperlink{sec:analytical_formula}{Section 2}  with the small discrepancy between the estimated and target values of the thermal diffusivity explained by the various numerical approximations~used.

\begin{table*}
\hypertarget{tab:results1}{}
\centering\small
\renewcommand{\arraystretch}{1.1}
\begin{tabular*}{1.0\textwidth}{@{\extracolsep{\fill}}llll}
\hline
& Disc-shaped slab ($d = 1$) & Cylindrical shell ($d = 2$) & Spherical shell ($d = 3$)\\
\hline
Outward Configuration &\\
$\widetilde{\alpha}$ [$\text{m}^{2}\text{s}^{-1}$] & \num{9.1766e-05} & \num{9.1766e-05} & \num{9.1765e-05}\\
$\varepsilon$ [$\%$] & \num{1.9997e-04} & \num{9.2837e-05} & \num{8.6366e-04}\\
Inward Configuration &\\
$\widetilde{\alpha}$ [$\text{m}^{2}\text{s}^{-1}$] & \num{9.1766e-05} & \num{9.1766e-05} & \num{9.1765e-05}\\
$\varepsilon$ [$\%$] & \num{1.9997e-04} & \num{9.2845e-05} & \num{8.6367e-04}\\
\hline
\end{tabular*}
\caption{Thermal diffusivity estimates, $\widetilde{\alpha}$, and corresponding signed relative percentages errors, $\varepsilon$, obtained from applying the formulas (\ref{eq:alpha_outward}) and (\ref{eq:alpha_inward}) and approximation (\ref{eq:trap_rule}), for the outward and inward configurations, respectively. All results correspond to the parameter values in (\ref{eq:parameters1})--(\ref{eq:parameters2}) and the ``perfect'' noise-free synthetic temperature rise data discussed in \protect\hyperlink{sec:perfect}{Section 3.2}.}
\end{table*}

\hypertarget{sec:noisy}{}
\subsection{Application to noisy synthetic data}
We now demonstrate that the thermal diffusivity formulas remain accurate when applied to noisy temperature rise data. This is achieved by perturbing the ``perfect data'' from the previous section using Gaussian noise yielding 
\begin{gather}
\label{eq:noisy_data}
\widetilde{T}_{i} = \begin{cases} T_{i}^{(0)} + \sigma z_{i}, & \text{for the inward configuration},\\
T_{i}^{(1)} + \sigma z_{i}, & \text{for the outward configuration}, \end{cases}
\end{gather}
for $i = 0,1,\hdots,N$, where $z_{i}\sim\mathcal{N}(0,1)$, $\sigma$ is the standard deviation controlling the level of noise and $T_{i}^{(0)}$ and $T_{i}^{(1)}$ are the numerical approximations to $T(\ell_{0},t_{i})$ and $T(\ell_{1},t_{i})$ as in the previous section. 

In \hyperlink{fig:one_realisation}{Figure 3}, we report results for two levels of noise, $\sigma = 0.02,0.05\,(^{\circ}\text{C})$, and one realisation of the random numbers $z_{1},\hdots,z_{N}$. Here we see that the theoretical temperature rise curves at $x=\ell_{0}$ and $x=\ell_{1}$, obtained by solving the governing heat flow model with $\alpha$ set equal to the estimated value $\widetilde{\alpha}$, provide an excellent visual fit to the noisy synthetic data. From these results it is clear that the dimension and configuration both impact the accuracy of the thermal diffusivity formulas in the presence of noise. For the outward configuration, it is clear from \hyperlink{fig:one_realisation}{Figures 3(a) and 3(c)} that the thermal diffusivity estimates are most accurate for the disc-shaped slab ($d=1$) and least accurate for the spherical shell ($d=3$) when comparing to the target value (\ref{eq:alpha_target}). The reason for this difference is due to the magnitude of $T_{\infty}$ (\ref{eq:Tinf_outward}). For the outward configuration, $T_{\infty}$ is largest for the disc-shaped slab and smallest for the spherical shell and therefore the added noise (\ref{eq:noisy_data}) for the spherical shell represents a larger relative perturbation of the perfect data that more greatly erodes the accuracy of the numerical integral approximation (\ref{eq:trap_rule}). A similar argument explains why the thermal diffusivity estimates are most accurate for the spherical shell ($d=3$) and least accurate for the disc-shaped slab ($d=1$) for the inward configuration (see \hyperlink{fig:one_realisation}{Figures 3(b) and 3(d)}). Note also that the same value of the thermal diffusivity is recorded for $d = 1$ in \hyperlink{fig:one_realisation}{Figures 3(a) and 3(b)} and \hyperlink{fig:one_realisation}{Figures 3(c) and 3(d)}, respectively, as distinguishing between the inwards/outwards direction is not required for the disc-shaped slab.

\def\figsize{0.44\textwidth}
\begin{figure}[t]
\hypertarget{fig:one_realisation}{}
\centering
\includegraphics[width=\figsize]{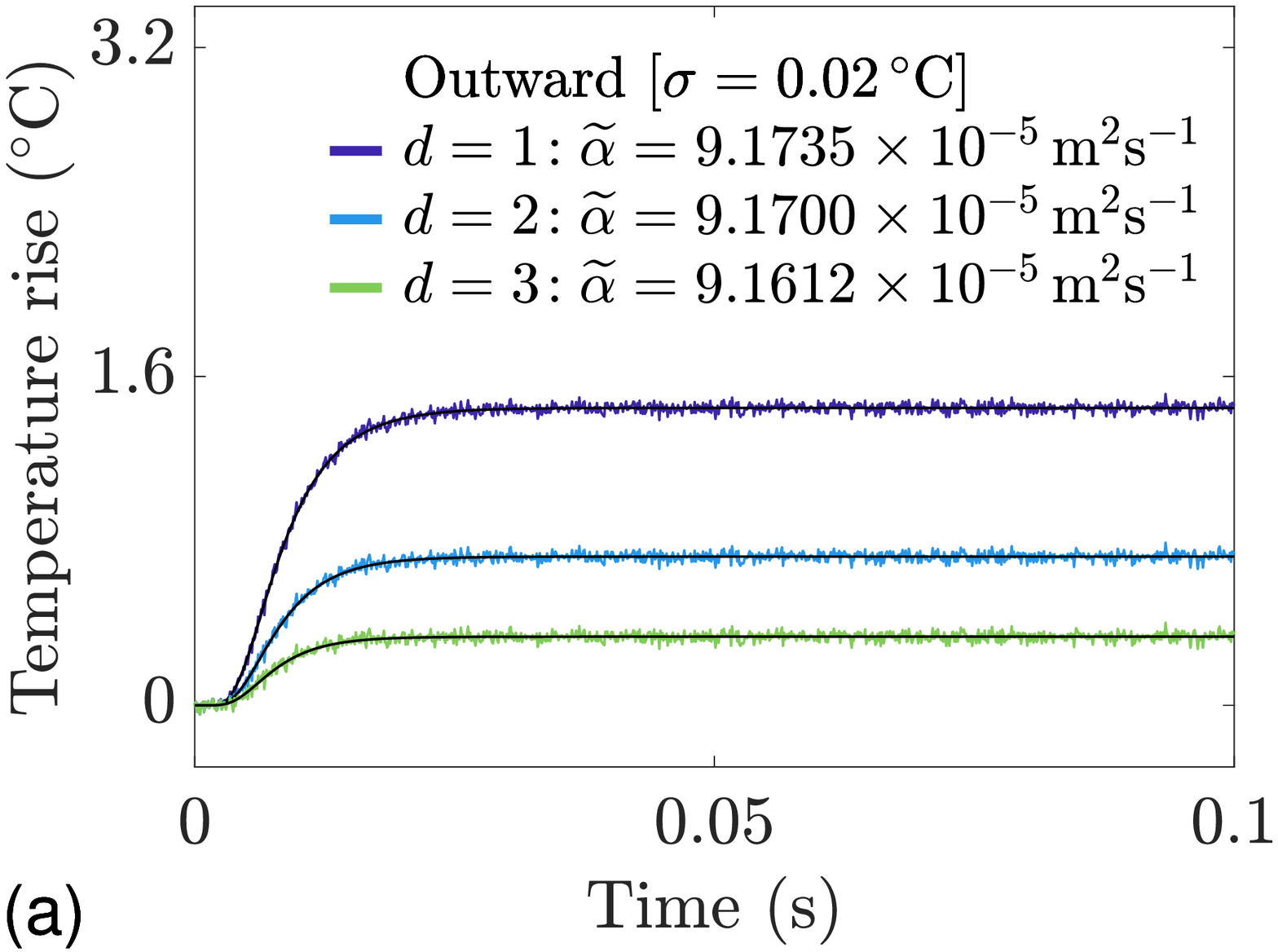}\hspace{0.05\textwidth}
\includegraphics[width=\figsize]{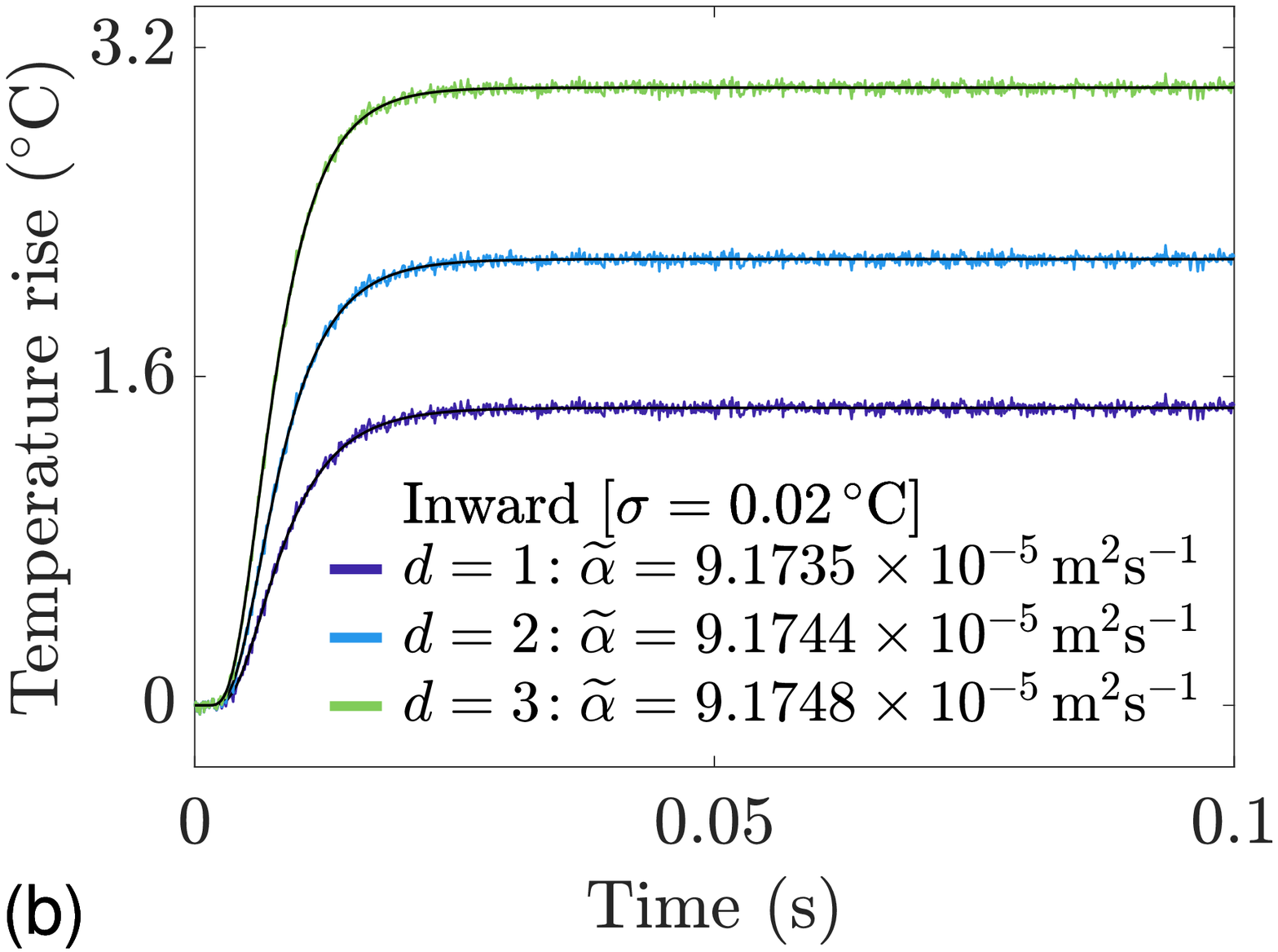}\\[0.02\textwidth]
\includegraphics[width=\figsize]{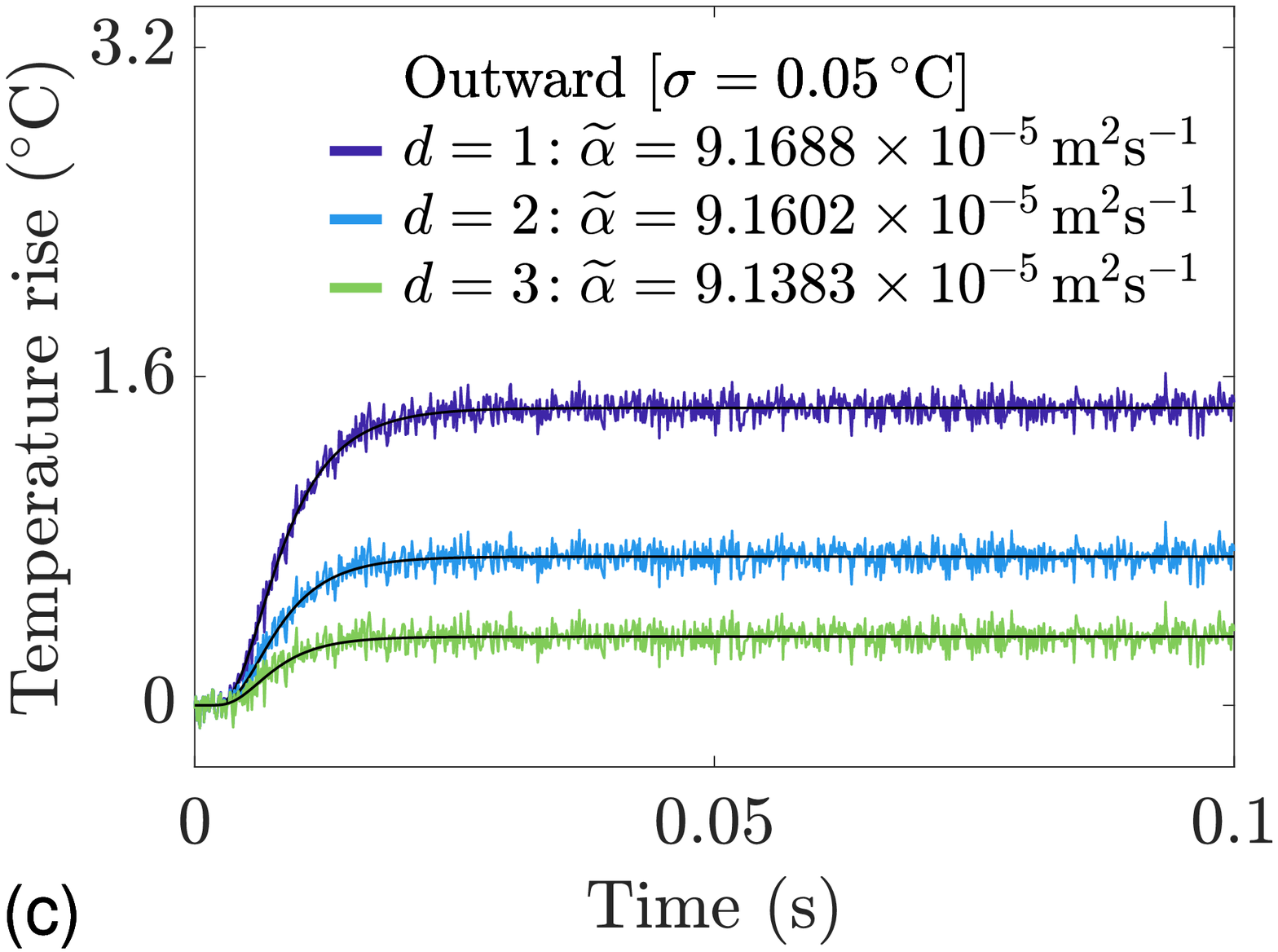}\hspace{0.05\textwidth}
\includegraphics[width=\figsize]{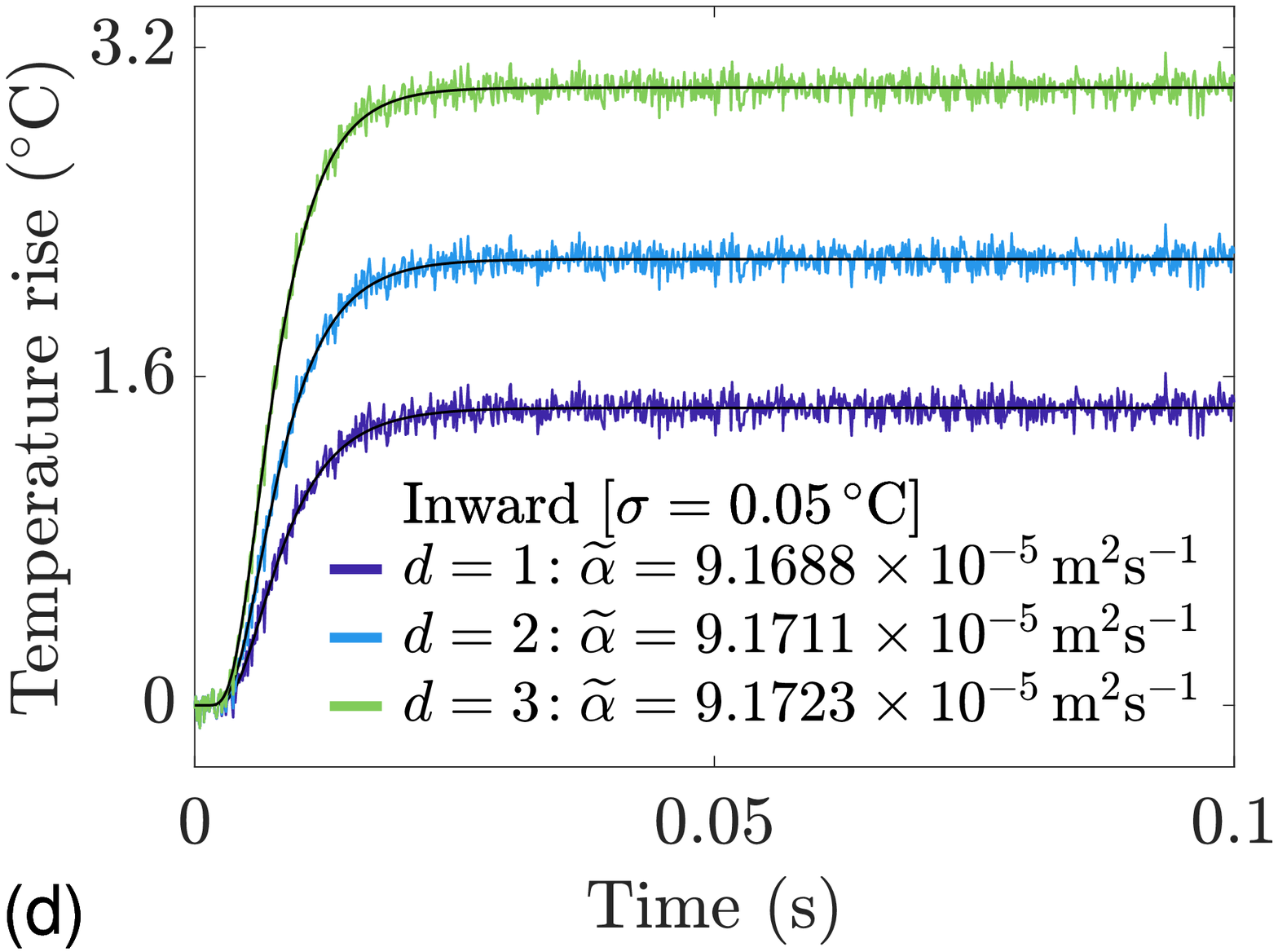}\\
\caption{Temperature rise curves for the (a) outward configuration with $\sigma = 0.02\,^{\circ}\text{C}$ (b) inward configuration with $\sigma = 0.02\,^{\circ}\text{C}$ (c) outward configuration with $\sigma = 0.05\,^{\circ}\text{C}$ and (d) inward configuration with $\sigma = 0.05\,^{\circ}\text{C}$. Each plot includes the calculated thermal diffusivity estimates, $\widetilde{\alpha}$, the noisy temperature rise data used to calculate $\widetilde{\alpha}$ (coloured lines) and the corresponding smooth temperature rise curves obtained by solving the governing heat flow model, equations (\ref{eq:pde_outward})--(\ref{eq:bcs_outward}) for the outward configuration and equations (\ref{eq:pde_outward}), (\ref{eq:ic_outward}) and (\ref{eq:bcs_inward}) for the inward configuration, with $\alpha = \widetilde{\alpha}$ (black lines). All results correspond to the parameter values in (\ref{eq:parameters1})--(\ref{eq:parameters2}) and one realisation of the random numbers $z_{1},\hdots,z_{N}$ as discussed in \protect\hyperlink{sec:noisy}{Section 3.3}.}
\end{figure}

\section{Conclusion}
\label{sec:conclusions}
We have revisited the rear-surface integral method for calculating the thermal diffusivity of solid materials, extending analytical formulas derived for disc-shaped slab samples with parallel front and rear-surfaces to the case of cylindrical-shell and spherical-shell shaped samples. Results for noise-free synthetic temperature rise data verified the presented thermal diffusivity formulas for the cylindrical shell and spherical shell under both the inward and outward configurations. Results for noisy synthetic temperature rise data obtained using additive Gaussian noise demonstrated that the outward configuration formula is most accurate for the disc-shaped slab and least accurate for the spherical shell (for noise with fixed standard deviation), while the opposite is true for the inward configuration formula. Both formulas are valid under the assumptions of a homogeneous sample and zero heat loss to the external environment. Potential directions for future work could therefore include accommodating composite layered samples and/or heat loss in the formulas.

\section*{Acknowledgements}
This work was partially supported by the Australian Mathematical Sciences Institute (AMSI) who provided LPF with a scholarship to undertake this research over the 2021-2022 Australian summer. We also wish to thank two anonymous reviewers for their helpful comments that improved the quality of the paper.

\bibliographystyle{model1-num-names}
\bibliography{references}

\begin{thebibliography}{25}
\expandafter\ifx\csname natexlab\endcsname\relax\def\natexlab#1{#1}\fi
\providecommand{\url}[1]{\texttt{#1}}
\providecommand{\href}[2]{#2}
\providecommand{\path}[1]{#1}
\providecommand{\DOIprefix}{doi:}
\providecommand{\ArXivprefix}{arXiv:}
\providecommand{\URLprefix}{URL: }
\providecommand{\Pubmedprefix}{pmid:}
\providecommand{\doi}[1]{\href{http://dx.doi.org/#1}{\path{#1}}}
\providecommand{\Pubmed}[1]{\href{pmid:#1}{\path{#1}}}
\providecommand{\bibinfo}[2]{#2}
\ifx\xfnm\relax \def\xfnm[#1]{\unskip,\space#1}\fi
%Type = Article
\bibitem[{Parker et~al.(1961)Parker, Jenkins, Butler, and Abbott}]{parker_1961}
\bibinfo{author}{W.~J. Parker}, \bibinfo{author}{R.~J. Jenkins},
  \bibinfo{author}{C.~P. Butler}, \bibinfo{author}{G.~L. Abbott},
\newblock \bibinfo{title}{Flash method of determining thermal diffusivity, heat
  capacity, and thermal conductivity},
\newblock \bibinfo{journal}{Journal of Applied Physics} \bibinfo{volume}{32}
  (\bibinfo{year}{1961}) \bibinfo{pages}{1679--1684}.
%Type = Article
\bibitem[{Cape and Lehman(1963)}]{cape_1963}
\bibinfo{author}{J.~A. Cape}, \bibinfo{author}{G.~W. Lehman},
\newblock \bibinfo{title}{Temperature and finite pulse-time effects in the
  flash method for measuring thermal diffusivity},
\newblock \bibinfo{journal}{Journal of Applied Physics} \bibinfo{volume}{34}
  (\bibinfo{year}{1963}) \bibinfo{pages}{1909--1913}.
%Type = Article
\bibitem[{Azumi and Takahashi(1981)}]{azumi_1981}
\bibinfo{author}{T.~Azumi}, \bibinfo{author}{Y.~Takahashi},
\newblock \bibinfo{title}{Novel finite pulse-wide correction in flash thermal
  diffusivity measurement},
\newblock \bibinfo{journal}{Review of Scientific Instruments}
  \bibinfo{volume}{52} (\bibinfo{year}{1981}) \bibinfo{pages}{1411--1413}.
%Type = Article
\bibitem[{Carr and Wood(2019)}]{carr_2019d}
\bibinfo{author}{E.~J. Carr}, \bibinfo{author}{C.~J. Wood},
\newblock \bibinfo{title}{Rear-surface integral method for calculating thermal
  diffusivity: Finite pulse time correction and two-layer samples},
\newblock \bibinfo{journal}{International Journal of Heat and Mass Transfer}
  \bibinfo{volume}{144} (\bibinfo{year}{2019}) \bibinfo{pages}{118609}.
%Type = Article
\bibitem[{Parker and Jenkins(1962)}]{parker_1962}
\bibinfo{author}{W.~J. Parker}, \bibinfo{author}{R.~J. Jenkins},
\newblock \bibinfo{title}{Thermal conductivity measurements on bismuth
  telluride in the presence of a 2 {MeV} electron beam},
\newblock \bibinfo{journal}{Advanced Energy Conversion} \bibinfo{volume}{2}
  (\bibinfo{year}{1962}) \bibinfo{pages}{87--103}.
%Type = Article
\bibitem[{Chen et~al.(2010)Chen, Limarga, and Clarke}]{chen_2010}
\bibinfo{author}{L.~Chen}, \bibinfo{author}{A.~M. Limarga},
  \bibinfo{author}{D.~R. Clarke},
\newblock \bibinfo{title}{A new data reduction method for pulse diffusivity
  measurements on coated samples},
\newblock \bibinfo{journal}{Computational Materials Science}
  \bibinfo{volume}{50} (\bibinfo{year}{2010}) \bibinfo{pages}{77--82}.
%Type = Article
\bibitem[{Carr(2019)}]{carr_2019b}
\bibinfo{author}{E.~J. Carr},
\newblock \bibinfo{title}{Rear-surface integral method for calculating thermal
  diffusivity from laser flash experiments},
\newblock \bibinfo{journal}{Chemical Engineering Science} \bibinfo{volume}{199}
  (\bibinfo{year}{2019}) \bibinfo{pages}{546--551}.
%Type = Article
\bibitem[{Larson and Koyama(1968)}]{larson_1968}
\bibinfo{author}{K.~B. Larson}, \bibinfo{author}{K.~Koyama},
\newblock \bibinfo{title}{Measurement by the flash method of thermal
  diffusivity, heat capacity, and thermal conductivity in two-layer composite
  samples},
\newblock \bibinfo{journal}{Journal of Applied Physics} \bibinfo{volume}{39}
  (\bibinfo{year}{1968}) \bibinfo{pages}{4408}.
%Type = Article
\bibitem[{Cz{\'e}l et~al.(2013)Cz{\'e}l, Woodbury, Woolley, and
  Gr{\'o}f}]{czel_2013}
\bibinfo{author}{B.~Cz{\'e}l}, \bibinfo{author}{K.~A. Woodbury},
  \bibinfo{author}{J.~Woolley}, \bibinfo{author}{G.~Gr{\'o}f},
\newblock \bibinfo{title}{Analysis of parameter estimation possibilities of the
  thermal contact resistance using the laser flash method with two-layer
  specimens},
\newblock \bibinfo{journal}{International Journal of Thermophysics}
  \bibinfo{volume}{34} (\bibinfo{year}{2013}) \bibinfo{pages}{1993--2008}.
%Type = Article
\bibitem[{Milo\v{s}evi\'{c} and Raynaud(2004)}]{milosevic_2004}
\bibinfo{author}{N.~D. Milo\v{s}evi\'{c}}, \bibinfo{author}{M.~Raynaud},
\newblock \bibinfo{title}{Analytical solution of transient heat conduction in a
  two-layer anisotropic cylindrical slab excited superficially by a short laser
  pulse},
\newblock \bibinfo{journal}{International Journal of Heat and Mass Transfer}
  \bibinfo{volume}{47} (\bibinfo{year}{2004}) \bibinfo{pages}{1627--1641}.
%Type = Article
\bibitem[{Takahashi et~al.(1988)Takahashi, Yamamoto, Ohsato, and
  Terai}]{takahashi_1988}
\bibinfo{author}{Y.~Takahashi}, \bibinfo{author}{K.~Yamamoto},
  \bibinfo{author}{T.~Ohsato}, \bibinfo{author}{T.~Terai},
\newblock \bibinfo{title}{Usefulness of logarithmic method in laser-flash
  technique for thermal diffusivity measurement},
\newblock \bibinfo{journal}{Proceedings of the 9th Japanese Symposium on
  Thermophysical Properties}  (\bibinfo{year}{1988}) \bibinfo{pages}{175--178}.
%Type = Article
\bibitem[{Thermitus and Laurent(1997)}]{thermitus_1997}
\bibinfo{author}{M.-A. Thermitus}, \bibinfo{author}{M.~Laurent},
\newblock \bibinfo{title}{New logarithmic technique in the flash method},
\newblock \bibinfo{journal}{International Journal of Heat and Mass Transfer}
  \bibinfo{volume}{40} (\bibinfo{year}{1997}) \bibinfo{pages}{4183--4190}.
%Type = Article
\bibitem[{Chihab et~al.(2020)Chihab, Garoum, and Laaroussi}]{chihab_2020}
\bibinfo{author}{Y.~Chihab}, \bibinfo{author}{M.~Garoum},
  \bibinfo{author}{N.~Laaroussi},
\newblock \bibinfo{title}{A new efficient formula for the thermal diffusivity
  estimation from the flash method taking into account heat losses in rear and
  front faces},
\newblock \bibinfo{journal}{International Journal of Thermophysics}
  \bibinfo{volume}{41} (\bibinfo{year}{2020}) \bibinfo{pages}{118}.
%Type = Article
\bibitem[{Nishi et~al.(2020)Nishi, Azuma, and Ohta}]{nishi_2020}
\bibinfo{author}{T.~Nishi}, \bibinfo{author}{N.~Azuma},
  \bibinfo{author}{H.~Ohta},
\newblock \bibinfo{title}{Effect of radiative heat loss on thermal diffusivity
  evaluated using normalized logarithmic method in laser flash technique},
\newblock \bibinfo{journal}{High Temperature Materials and Processes}
  \bibinfo{volume}{39} (\bibinfo{year}{2020}) \bibinfo{pages}{390--394}.
%Type = Article
\bibitem[{Clark and Taylor(1975)}]{clark_1975}
\bibinfo{author}{L.~M. Clark}, \bibinfo{author}{R.~E. Taylor},
\newblock \bibinfo{title}{Radiation loss in the flash method for thermal
  diffusivity},
\newblock \bibinfo{journal}{Journal of Applied Physics} \bibinfo{volume}{46}
  (\bibinfo{year}{1975}) \bibinfo{pages}{714--719}.
%Type = Article
\bibitem[{Gosset and Colin(2002)}]{gosset_2002}
\bibinfo{author}{D.~Gosset}, \bibinfo{author}{M.~Colin},
\newblock \bibinfo{title}{Improvement of the mathematical modelling of flash
  measurements},
\newblock \bibinfo{journal}{High Temperatures - High Pressures}
  \bibinfo{volume}{34} (\bibinfo{year}{2002}) \bibinfo{pages}{265--280}.
%Type = Article
\bibitem[{Voz{\'a}r and Hohenauer(2003)}]{vozar_2003}
\bibinfo{author}{L.~Voz{\'a}r}, \bibinfo{author}{W.~Hohenauer},
\newblock \bibinfo{title}{Flash method of measuring the thermal diffusivity.
  {A} review},
\newblock \bibinfo{journal}{High Temperatures - High Pressures}
  \bibinfo{volume}{35--36} (\bibinfo{year}{2003}) \bibinfo{pages}{253--264}.
%Type = Article
\bibitem[{Degiovanni(1986)}]{degiovanni_1986}
\bibinfo{author}{A.~Degiovanni},
\newblock \bibinfo{title}{A new technique for identifying thermal diffusivity
  for the “flash” method (in french)},
\newblock \bibinfo{journal}{Revue de Physique Appliqu\'ee} \bibinfo{volume}{21}
  (\bibinfo{year}{1986}) \bibinfo{pages}{229--237}.
%Type = Article
\bibitem[{Gembarovic et~al.(1990)Gembarovic, Vozar, and
  Majernik}]{gembarovic_1990}
\bibinfo{author}{J.~Gembarovic}, \bibinfo{author}{L.~Vozar},
  \bibinfo{author}{V.~Majernik},
\newblock \bibinfo{title}{Using the least square method for data reduction in
  the flash method},
\newblock \bibinfo{journal}{International Journal of Heat and Mass Transfer}
  \bibinfo{volume}{33} (\bibinfo{year}{1990}) \bibinfo{pages}{1563--1565}.
%Type = Article
\bibitem[{Rassy et~al.(2020)Rassy, Billaud, and Saury}]{el-rassy_2020}
\bibinfo{author}{E.~E. Rassy}, \bibinfo{author}{Y.~Billaud},
  \bibinfo{author}{D.~Saury},
\newblock \bibinfo{title}{Flash method experiment design for the thermal
  characterization of orthotropic materials},
\newblock \bibinfo{journal}{Measurement Science and Technology}
  \bibinfo{volume}{31} (\bibinfo{year}{2020}) \bibinfo{pages}{085901}.
%Type = Article
\bibitem[{Lunev and Heymer(2020)}]{lunev_2020}
\bibinfo{author}{A.~Lunev}, \bibinfo{author}{R.~Heymer},
\newblock \bibinfo{title}{Decreasing the uncertainty of classical laser flash
  analysis using numerical algorithms robust to noise and systematic errors},
\newblock \bibinfo{journal}{Review of Scientific Instruments}
  \bibinfo{volume}{91} (\bibinfo{year}{2020}) \bibinfo{pages}{064902}.
%Type = Article
\bibitem[{Milo\v{s}evi\'{c}(2019)}]{milosevic_2019}
\bibinfo{author}{N.~D. Milo\v{s}evi\'{c}},
\newblock \bibinfo{title}{Analytical solution of a generalized two-dimensional
  model for laser-flash thermal diffusivity measurements},
\newblock \bibinfo{journal}{Journal of Thermophysics and Heat Transfer}
  \bibinfo{volume}{33} (\bibinfo{year}{2019}) \bibinfo{pages}{300--308}.
%Type = Article
\bibitem[{Lamien et~al.(2019)Lamien, Maux, Courtois, Pierre, Carin, Masson,
  Orlande, and Paillard}]{lamien_2019}
\bibinfo{author}{B.~Lamien}, \bibinfo{author}{D.~L. Maux},
  \bibinfo{author}{M.~Courtois}, \bibinfo{author}{T.~Pierre},
  \bibinfo{author}{M.~Carin}, \bibinfo{author}{P.~L. Masson},
  \bibinfo{author}{H.~R.~B. Orlande}, \bibinfo{author}{P.~Paillard},
\newblock \bibinfo{title}{A {Bayesian} approach for the estimation of the
  thermal diffusivity of aerodynamically levitated solid metals at high
  temperatures},
\newblock \bibinfo{journal}{High Temperature Materials and Processes}
  \bibinfo{volume}{141} (\bibinfo{year}{2019}) \bibinfo{pages}{265--281}.
%Type = Article
\bibitem[{Yan et~al.(2022)Yan, Li, Wang, Fa, and Zhang}]{yan_2022}
\bibinfo{author}{B.~Yan}, \bibinfo{author}{B.~Li}, \bibinfo{author}{X.~Wang},
  \bibinfo{author}{T.~Fa}, \bibinfo{author}{P.~Zhang},
\newblock \bibinfo{title}{Measuring thermal conductivity of materials at room
  temperature in atmosphere by using a continuous-wave laser and neural network
  model},
\newblock \bibinfo{journal}{International Journal of Heat and Mass Transfer}
  \bibinfo{volume}{189} (\bibinfo{year}{2022}) \bibinfo{pages}{122704}.
%Type = Article
\bibitem[{Baba(2009)}]{baba_2009}
\bibinfo{author}{T.~Baba},
\newblock \bibinfo{title}{Analysis of one-dimensional heat diffusion after
  light pulse heating by the response function method},
\newblock \bibinfo{journal}{Japanese Journal of Applied Physics}
  \bibinfo{volume}{48} (\bibinfo{year}{2009}) \bibinfo{pages}{05EB04}.

\end{thebibliography}

\end{document}